\documentclass[a4paper,12pt]{article}
\usepackage{epsfig}
\usepackage{rotating}
\usepackage{cite}
\def\21{$SU(2) \otimes U(1) $}

\newcommand{\Sol}  {\textsc{sol}}
\newcommand{\Atm}  {\textsc{atm}}
\newcommand{\chooz}  {\textsc{chooz}}

\newcommand{\Dms}  {\Delta m^2_\Sol}
\newcommand{\Dma}  {\Delta m^2_\Atm}

\newcommand{\AddrAHEP}{%
  AHEP Group, Institut de F\'{\i}sica Corpuscular --
  C.S.I.C./Universitat de Val{\`e}ncia \\
  Edificio de Institutos de Paterna, Apartado 22085,
  E--46071 Val{\`e}ncia, Spain}

\begin{document} 
\begin{titlepage} 
\begin{flushright}
hep-ph/0304141 \\ 
IFIC/03-10\\
\end{flushright} 
\vspace*{3mm} 
\begin{center}  

  
  \textbf{\large Reconstructing Neutrino Properties from Collider
    Experiments in a Higgs Triplet Neutrino Mass Model}\\[10mm]

{D. Aristiz\'abal Sierra${}^1$, M. Hirsch${}^2$, J. W. F. Valle${}^2$ and 
A. Villanova del Moral${}^2$} \vspace{0.3cm}\\ 

$^1$ Departamento de F{\'\i}sica,  Universidad de Antioquia \\
A.A. {\it 1226} \ Medell{\'\i}n,  \ Colombia 

$^2$ \AddrAHEP

\end{center}

\begin{abstract} 
  We extend the minimal supersymmetric standard model with bilinear
  R-parity violation to include a pair of Higgs triplet superfields.
  The neutral components of the Higgs triplets develop small vacuum
  expectation values (VEVs) quadratic in the bilinear R-parity
  breaking parameters. In this scheme the atmospheric neutrino mass
  scale arises from bilinear R-parity breaking while for reasonable  
  values of parameters the solar
  neutrino mass scale is generated from the small Higgs triplet VEVs.
  We calculate neutrino masses and mixing angles in this model and
  show how the model can be tested at future colliders. The branching
  ratios of the doubly charged triplet decays are related to the solar
  neutrino angle via a simple formula.
\end{abstract} 
 
\end{titlepage}

\newpage

\setcounter{page}{1}

\section{Introduction}

Ever since its proposal as a way to generate neutrino masses, either
on its own or in the context of effective seesaw
schemes~\cite{Schechter:1980gr}, many variants of the triplet model
have been considered\footnote{Here the triplet majoron will not be 
discussed because it is ruled out by the measurement of the invisible 
Z decay width at LEP.}.  In fact the triplet is a common ingredient
in the formulation of seesaw schemes with
gauged~\cite{Gell-Mann:1980vs,Yanagida:1979,Mohapatra:1981yp}, or
ungauged B-L symmetry~\cite{Chikashige:1981ui,Schechter:1982cv}.  They
are also a characteristic feature in several left-right symmetric
models \cite{Pati:1974yy}, required in order to reduce the extended
gauge structure down to the minimal \21. For studies of the
phenomenology of left-right symmetric models, both supersymmetric and
non-supersymmetric see ref.~\cite{Deshpande:1991ip} and references
therein.

Recent experiments, including the recently published first results of
KamLAND \cite{Eguchi:2002dm}, have confirmed the LMA-MSW oscillation
solution to the solar neutrino problem~\cite{Pakvasa:2003zv}. Together
with the earlier discoveries in atmospheric neutrinos
\cite{fukuda:1998mi}, one can now be fairly confident that all
neutrino flavours mix and that at least two non-zero neutrino masses
exist. This impressive progress has brought the quest for an understanding 
of the smallness of neutrino masses to the center of attention in particle
physics.

In the standard model neutrinos are massless. One of the best-known
mechanisms to generate small (Majorana) neutrino masses is the seesaw
mechanism.  Elegant as it may be, the seesaw mechanism is not the only
theoretically interesting approach to neutrino masses.  More
interesting from a phenomenological point of view are models in which
the neutrino masses are generated at the electroweak scale.  Models of
this kind include, for example, variants of the seesaw scheme at the 
electro-weak scale~\cite{Mohapatra:1986bd,Gonzalez-Garcia:1989rw}, radiative
models of neutrino mass~\cite{Zee:1980ai,Babu:1988ki} and
supersymmetry with R-parity
violation~\cite{hall:1984id,Santamaria:zm,banks:1995by,kaplan:1999ds,akeroyd:1998iq,choi:2000bm}.

In this work we further explore the triplet approach to neutrino mass,
by combining it with the idea of bilinear R-parity violating (BRpV)
supersymmetry.  The later might arise as the effective description of
models with spontaneously broken R-parity~\cite{Masiero:1990uj,Abreu:2000mm}.
Alternatively, the
required smallness of the bilinear R parity violating parameters may
result from some suitable family symmetry~\cite{Mira:2000gg}, thus
providing a common solution \cite{Nilles:1997ij} to the so-called
mu-problem~\cite{Giudice:1988yz} and the neutrino anomalies.  

At the superpotential level lepton
number violation resides only in the explicit BRpV terms, but after
electro-weak symmetry breaking takes place, this induces a naturally
small VEV for the neutral component of the Higgs triplets, suppressed
by two powers of the BRpV parameters~\cite{Santamaria:zm}, as
emphasized recently by Ernest Ma~\cite{Ma:2002xt}~\footnote{This
  corresponds to a type II weak-scale supersymmetric seesaw scheme,
  where an effective triplet VEV is induced from the exchange of
  scalar bosons.  This should be contrasted with the type-I mechanism
  which produces the effective neutrino mass from the exchange of
  neutralinos. Both types, however, come out proportional to the square 
  of RPV parameters, as seen in eq.~(\ref{eq:VEVs}) and eq.~(\ref{MBeff}) 
  below.  }.

The result is a hybrid scheme for the neutrino
masses.  An attractive and natural possibility is that the atmospheric
mass scale arises from bilinear R-parity breaking, while the solar
mass scale is generated by the small Higgs triplet VEVs. The model is
theoretically simple, since only tree-level physics is required to
explain current neutrino data, and has the advantage of being directly
testable in the next generation of colliders.

The presence of Higgs triplets would induce lepton flavor violating
decays of muons and taus.  A number of low-energy constraints on Higgs
triplet couplings have been derived from the non-observation of such
processes~\cite{Swartz:1989qz}. Although the model we are considering
is different from the standard left-right models, many constraints
apply equally well to our case.  The most important bounds for our
model are from the experimental upper limits on $\mu \to 3 e$ and $\mu
\to e \gamma$, which can be expressed as \cite{Swartz:1989qz},

\begin{equation}
h_{e\mu} h_{ee} < 3.2 \times 10^{-7} \frac{M_{\Delta^{--}_u}^2}
                                          {(\rm 100 \hskip1mm GeV)^2}
\label{mu3e}
\end{equation}
from $\mu \to  3 e$ and 
\begin{equation}
h_{e\mu} h_{\mu\mu} < 2 \times 10^{-6} \frac{M_{\Delta^{--}_u}^2}
                                          {(\rm 100 \hskip1mm GeV)^2}
\label{muegamma}
\end{equation}
from $\mu \to e \gamma$. Here, $h_{ij}$ are the Yukawa couplings of
the triplet to the leptons and $M_{\Delta^{--}_u}$ is the triplet mass.
Note that these bounds are on products of two couplings.  Limits exist
\cite{Swartz:1989qz} on individual Yukawa couplings but are weaker by
several orders of magnitude.

This paper is organized as follows.  In the next section we will
present the model, discussing the Higgs potential as well as the mass
matrix describing the neutrino-neutralino sector. In
Sec.~\ref{sec:phenomenology} we will then turn to the phenomenology of
the model. Neutrino masses and mixings are calculated, with emphasis
on the solar neutrino data.  Finally production and decays of the
Higgs bosons of the model are discussed.  In Sec.~\ref{sec:impl-accel}
we point out that there is a relation between various branching ratios
of the decay of the doubly charged component of the scalar triplet and
the solar angle, which will allow to test the validity of the model in
a future accelerator.  We will then close with a short summary.

\section{Model}

\subsection{Superpotential and scalar sector}

The model to be presented below is the supersymmetric extension of
the original triplet model of neutrino mass~\cite{Schechter:1980gr}
in which the simplest form of R-parity violation is assumed. The model is
defined by the particle content of the Minimal Supersymmetric
extension of Standard Model (MSSM) augmented by a pair of Higgs
triplet superfields:
\begin{equation}
\widehat{\Delta}_u=
\left(\begin{array}{l}
\widehat{\Delta}^{++}_u\\
\widehat{\Delta}^{+}_u\\
\widehat{\Delta}^{0}_u
\end{array}\right)
\qquad
\widehat{\Delta}_d=
\left(\begin{array}{l}
\widehat{\Delta}^{0}_d\\
\widehat{\Delta}^{-}_d\\
\widehat{\Delta}^{--}_d
\end{array}\right)
\end{equation}
with hypercharges $Y=+2$ and $Y=-2$ and lepton number $L=-2$ and
$L=+2$ respectively. Note that although only one Higgs triplet
superfield is necessary to provide the neutrinos with appropriate
masses, two triplets are needed to avoid the triangle gauge anomaly.
Apart from the new Higgs triplet superfields, we include R-parity
violation in three generations, by adding bilinear lepton number
violating superpotential
terms~\cite{Diaz:1998xc},~\cite{hall:1984id,banks:1995by,kaplan:1999ds,akeroyd:1998iq,choi:2000bm},
~\cite{Hirsch:2000ef,Diaz:2003as}. The superpotential of this model is
then given by a sum of three terms,
\begin{equation} \label{eq:superpotential}
W=W_{MSSM}+W_{BRpV}+W_{\Delta}
\end{equation}
where
\begin{equation}\label{eq:MSSM-superpotential}
W_{MSSM}=-\mu \widehat{H}_u\widehat{H}_d+h_U^{ij}
\widehat{H}_u\widehat{Q}_i\widehat{u}^c_j+h_D^{ij}
\widehat{H}_d\widehat{Q}_i\widehat{d}^c_j+h_E^{ij}
\widehat{H}_d\widehat{L}_i\widehat{e}^c_j
\end{equation}
is the superpotential of the MSSM,
\begin{equation}\label{eq:BRpV-superpotential}
W_{BRpV}=\epsilon_i\widehat{L}_i\widehat{H}_u
\end{equation}
is the usual bilinear R-parity violating term, while  
\begin{equation}\label{eq:triplet-superpotential}
W_{\Delta}=\mu_{\Delta}\widehat{\Delta}_u\widehat{\Delta}_d+h_{ij}
\widehat{L}_i\widehat{L}_j\widehat{\Delta}_u
\end{equation}
includes the terms that determine the interactions and masses of the
new Higgs triplet superfields.
The Yukawa couplings $h_U$, $h_D$, $h_E$ and $h$ are $3\times3$ matrices 
in generation space and $\mu$, $\epsilon_i$ ($i=1,\ldots,3$) and $M$ are 
parameters with units of mass.

The scalar potential along neutral directions is a sum of two terms
\begin{equation}\label{eq:V}
V=V_{SUSY}+V_{soft}
\end{equation}
where
\begin{equation}\begin{array}{ccl}
V_{SUSY} & = & \left|\mu_{\Delta}\Delta_u^0\right|^2+\left|h^{ij}
\widetilde{\nu}_i\widetilde{\nu}_j+\mu_{\Delta}\Delta_d^0\right|^2+
\left|\mu H_u^0\right|^2+\\
 & & \\
 & & +\frac{1}{8}(g^2+g^{\prime 2})\left(\left|H_d^0\right|^2-
\left|H_u^0\right|^2+\sum_i\left|\widetilde{\nu}_i\right|^2+2
\left|\Delta_d^0\right|^2-2\left|\Delta_u^0\right|^2\right)^2+\\
 & & \\
 & & +\sum_i\left|2h^{ij}\widetilde{\nu}_j\Delta_u^0
+H_u^0\epsilon^i\right|^2+\left|-\mu H_d^0+\epsilon^i\widetilde{\nu}_i\right|^2
\end{array}\end{equation}
is the neutral part of the supersymmetric scalar potential and
\begin{equation}\label{eq:V-soft}\begin{array}{ccl}
V_{soft} & = & \left[A^{ij}h^{ij}\widetilde{\nu}_i\widetilde{\nu}_j
\Delta_u^0-B\mu H_d^0H_u^0+B^i\epsilon^i\widetilde{\nu}_iH_u^0+B_{\Delta}\mu_{\Delta}
\Delta_u^0\Delta_d^0+c.c.\right]+\\
 & & \\
 & & +M_L^2\sum_i\left|\widetilde{\nu}_i\right|^2+
M_{H_d}^2\left|H_d^0\right|^2+M_{H_u}^2\left|H_u^0\right|^2+\\
 & & \\
 & & +M_{\Delta_u}^2\left|\Delta_u^0\right|^2
+M_{\Delta_d}^2\left|\Delta_d^0\right|^2
\end{array}\end{equation}
is the soft supersymmetry breaking scalar potential, also along
neutral directions.  The expression in eq.~(\ref{eq:V-soft}) contains
terms which are linear in the neutral Higgs bosons $H_u^0$, $H_d^0$,
$\Delta_u^0$, $\Delta_d^0$ and the scalar neutrinos
$\widetilde{\nu}_i$ ($i=1,\ldots,3$). The presence of bilinear RPV
terms in the superpotential leads to non-zero sneutrino vacuum
expectation values.

The vacuum expectation values can be determined by minimizing the scalar 
potential in eq.~(\ref{eq:V}). These stationary conditions of the scalar 
potential are the so-called tadpole equations: 
\begin{equation} \label{eq:tadpole-equations}
\begin{array}{lcl}
t_u & = & -\mu Bv_d+(\mu^2+M^2_{H_u})v_u+B^i\epsilon^iv_i
+\sqrt{2}\epsilon^ih_{ij}v^j\langle\Delta^0_u\rangle+ \\
 & & \\
 & & +\epsilon^i\epsilon_iv_u-v_u\mathcal{D} \\
 & & \\
t_d & = & (\mu^2+M^2_{H_d})v_d-\mu(Bv_u+\epsilon^iv_i)+v_d\mathcal{D} \\
 & & \\
t^i & = & \epsilon^i(-\mu v_d+B^iv_u+\epsilon^jv_j)
+M^2_Lv^i+\sqrt{2}h^{ij}\epsilon_jv_u\langle\Delta^0_u\rangle+ \\
 & & \\
 & & +\sqrt{2}A^{ij}h^{ij}v_j\langle\Delta^0_u\rangle
+\sqrt{2}\mu_{\Delta}h^{ij}v_j\langle\Delta^0_d\rangle+h^{ij}v_jh^{kl}v_kv_l+ \\
 & & \\
 & & +2h^{ij}h_{jk}v^k\langle\Delta^0_u\rangle^2+v^i\mathcal{D} \\
 & & \\
t_{\langle\Delta^0_u\rangle} & = & (\mu_{\Delta}^2+M^2_{\Delta_u})
\langle\Delta^0_u\rangle+v^ih_{ij}(2h^{jk}v_k
\langle\Delta^0_u\rangle+\sqrt{2}\epsilon^jv_u)+ \\
 & & \\
 & & +\frac{1}{\sqrt{2}}A^{ij}h^{ij}v_iv_j
+B_{\Delta}\mu_{\Delta}\langle\Delta^0_d\rangle-2\langle\Delta^0_u\rangle\mathcal{D} \\
 & & \\
t_{\langle\Delta^0_d\rangle} & = & (\mu_{\Delta}^2+M^2_{\Delta_d})
\langle\Delta^0_d\rangle+\frac{1}{\sqrt{2}}\mu_{\Delta}h^{ij}v_iv_j
+B_{\Delta}\mu_{\Delta}\langle\Delta^0_u\rangle+2\langle\Delta^0_d\rangle\mathcal{D}
\end{array}
\end{equation}
where the non-zero vacuum expectation values are defined as
\begin{equation}\label{eq:definition-VEVs}
v_u\equiv\langle H_u^0\rangle,\quad\quad v_d\equiv
\langle H_d^0\rangle,\quad\quad v_i\equiv
\langle\widetilde{\nu}_i\rangle\quad (i=,1\ldots,3) 
\end{equation}
and
\begin{equation}\label{eq:Dterm}
\mathcal{D}\equiv\frac{1}{8}(g^2+g'^2)(v^2_d-v^2_u
+\sum_iv^2_i+2\langle\Delta^0_d\rangle^2-2\langle\Delta^0_u\rangle^2)
\end{equation}

The basic idea of the model is contained in eqs.~(\ref{eq:tadpole-equations}) 
and (\ref{eq:Dterm}) and can be 
understood with the help of the following consideration. 
The VEVs of the scalar triplets violate lepton number by two 
units and, therefore, are expected to be small. If 
$\langle\Delta^0_u\rangle,\langle\Delta^0_d\rangle \ll v_u,v_d$, 

\begin{equation}\label{eq:DtermApp}
\mathcal{D}\simeq \mathcal{D}'\equiv\frac{1}{8}(g^2+g'^2)(v^2_d-v^2_u
+\sum_iv^2_i)
\end{equation}
and one can solve eqs.~(\ref{eq:tadpole-equations}) for 
$\langle\Delta^0_u\rangle$ and $\langle\Delta^0_d\rangle$,

\begin{equation} \label{eq:VEVs}
\begin{array}{lcl}
\langle\Delta^0_u\rangle & \simeq & 
\frac{1}{\sqrt{2}}h^{ij}\frac{v_iv_j[-A^{ij}(\mu_{\Delta}^2+M^2_{\Delta_d}+2\mathcal{D}')+B_{\Delta}\mu_{\Delta}^2]-2v_i\epsilon_jv_u(\mu_{\Delta}^2+M^2_{\Delta_d}+2\mathcal{D}')}
{(\mu_{\Delta}^2+M^2_{\Delta_u}-2\mathcal{D}')(\mu_{\Delta}^2+M^2_{\Delta_d}+2\mathcal{D}')-(B_{\Delta}\mu_{\Delta})^2}\\
 & & \\
\langle\Delta^0_d\rangle & \simeq & 
\frac{1}{\sqrt{2}}h^{ij}\frac{v_iv_j[-A^{ij}B_{\Delta}\mu_{\Delta}+\mu_{\Delta}(\mu_{\Delta}^2+M^2_{\Delta_u}-2\mathcal{D}')]-2v_i\epsilon_jv_uB_{\Delta}\mu_{\Delta}}
{(\mu_{\Delta}^2+M^2_{\Delta_u}-2\mathcal{D}')(\mu_{\Delta}^2+M^2_{\Delta_d}+2\mathcal{D}')-(B_{\Delta}\mu_{\Delta})^2}
\end{array}
\end{equation}
The Higgs triplet VEVs are quadratic in the RPV parameters, and thus
automatically small as long as $v_i \ll v_d,v_u$. The smallness of the
$v_i$, on the other hand, depends on the relative size of
$\epsilon_i/\mu$ which we assume to be sufficiently smaller than one.
Note also that the presence of neutral scalar lepton VEVs produces a
mixing between scalars, higgs bosons and scalar
neutrinos~\cite{akeroyd:1998iq}.

\subsection{Neutral Fermion Mixing}

The Lagrangian contains the following terms involving two neutral
fermions and the neutral scalars (Higgs bosons and scalar neutrinos)
\begin{equation}\begin{array}{ccl}
\mathcal{L} & \supset & \mu\widetilde{H}_d^0\widetilde{H}_u^0
-\epsilon^i\nu_i\widetilde{H}_u^0+h^{ij}\nu_i\nu_j\Delta_u^0
+h^{ij}\nu_i\widetilde{\nu}_j\widetilde{\Delta}_u^0
-\mu_{\Delta}\widetilde{\Delta}_u^0\widetilde{\Delta}_d^0+\\
 & & \\
 & & +\frac{1}{2}(M_1\lambda^{\prime}\lambda^{\prime}+M_2\lambda\lambda)+\\
 & & \\
 & & +\frac{i}{\sqrt{2}}g^{\prime}\lambda^{\prime}
(-\widetilde{\nu}^{i*}\nu_i-H_d^{0*}\widetilde{H}_d^0
+H_u^{0*}\widetilde{H}_u^0+2\Delta_u^{0*}\widetilde{\Delta}_u^0
-2\Delta_d^{0*}\widetilde{\Delta}_d^0)+\\
 & & \\
 & & +\frac{i}{\sqrt{2}}g\lambda^3(\widetilde{\nu}^{i*}\nu_i
+H_d^{0*}\widetilde{H}_d^0-H_u^{0*}\widetilde{H}_u^0
-2\Delta_u^{0*}\widetilde{\Delta}_u^0+2\Delta_d^{0*}\widetilde{\Delta}_d^0)
\end{array}\end{equation}
After the electroweak and R-parity symmetries break through non-zero
vacuum expectation values for the Higgs boson and sneutrinos, the
Lagrangian contains the following mass terms involving nine neutral
fermions:
\begin{equation}
\mathcal{L}\supset -\frac{1}{2}(\psi^0)^T\mathbf{M_N}(\psi^0)
\end{equation}
where the basis is
\begin{equation} \label{eq:basis}
(\psi^0)^T=(\nu_1,\nu_2,\nu_3,-i\lambda',-i\lambda^3,\widetilde{H}^0_d,
\widetilde{H}^0_u,\widetilde{\Delta}^0_d,\widetilde{\Delta}^0_u)
\end{equation}
and the neutral fermion mass matrix is a $9\times 9$ matrix of the form:
\begin{equation} \label{eq:neutral-fermion-mass-matrix}
\mathbf{M_N}=
\left(\begin{array}{lll}
\mathbf{M_{\nu}} & \mathbf{m}_{\mathbf{BRpV}} & 
  \mathbf{m}_{\mathbf{\nu\Delta}} \\
\\
\mathbf{m}^T_{\mathbf{BRpV}} & \mathbf{\mathcal{M}_{\chi^0}} & 
  \mathbf{m}_{\mathbf{\chi^0\Delta}} \\
\\
\mathbf{m}^T_{\mathbf{\nu\Delta}} & \mathbf{m}^T_{\mathbf{\chi^0\Delta}} & 
  \mathbf{M_{\Delta}}
\end{array} \right)
\end{equation}
where the symmetric matrix
\begin{equation} \label{eq:neutrino-mass-matrix}
\mathbf{M_{\nu }}=\sqrt{2}\langle \Delta^0_u\rangle  
\left(\begin{array}{ccc}
h_{11} & h_{12} & h_{13} \\
h_{12} & h_{22} & h_{23} \\
h_{13} & h_{23} & h_{33}
\end{array} \right)
\end{equation}
characterizes the direct contribution to the neutrino mass matrix and
\begin{equation} \label{eq:neutralino-mass-matrix}
\mathbf{\mathcal{M}_{\chi^0}}=
\left(\begin{array}{cccc}
M_1 & 0 & -\frac{1}{2}g'v_d & \frac{1}{2}g'v_u \\
0 & M_2 & \frac{1}{2}gv_d & -\frac{1}{2}gv_u \\
-\frac{1}{2}g'v_d & \frac{1}{2}gv_d & 0 & -\mu \\
\frac{1}{2}g'v_u & -\frac{1}{2}gv_u & -\mu & 0 
\end{array} \right)
\end{equation}
denotes the standard MSSM neutralino mass matrix. The matrix
$\mathbf{m_{BRpV}}$ is given as
\begin{equation} \label{eq:BRpV-mass-matrix}
\mathbf{m_{BRpV}}=
\left(\begin{array}{cccc}
-\frac{1}{2}g'v_1 & \frac{1}{2}gv_1 & 0 & \epsilon_1 \\
-\frac{1}{2}g'v_2 & \frac{1}{2}gv_2 & 0 & \epsilon_2 \\
-\frac{1}{2}g'v_3 & \frac{1}{2}gv_3 & 0 & \epsilon_3
\end{array} \right)
\end{equation}
and characterizes the bilinear R-parity Violation, while
\begin{equation} \label{eq:neutrino-triplet-mass-matrix}
\mathbf{m_{\nu\Delta}}=
\left(\begin{array}{cc}
0 & \frac{1}{\sqrt{2}}h^{1j}v_j \\
0 & \frac{1}{\sqrt{2}}h^{2j}v_j \\
0 & \frac{1}{\sqrt{2}}h^{3j}v_j
\end{array} \right)
\end{equation}
is the neutrino-Higgs triplet mixing mass matrix. On the other hand,
\begin{equation} \label{eq:neutralino-triplet-mass-matrix}
\mathbf{m_{\chi^0\Delta}}=
\left(\begin{array}{cc}
-g'\langle \Delta^0_d\rangle & g'\langle \Delta^0_u\rangle \\
g\langle\Delta^0_d\rangle & -g\langle \Delta^0_u\rangle \\
0 & 0 \\
0 & 0
\end{array} \right)
\end{equation}
is the neutralino-Higgs triplet mixing mass matrix and
\begin{equation} \label{eq:triplet-mass-matrix}
\mathbf{M_{\Delta}}=
\left(\begin{array}{cc}
0 & \mu_{\Delta} \\
\mu_{\Delta} & 0
\end{array} \right)
\end{equation}
is the Higgs triplet mass matrix arising from the first term in
eq.~(\ref{eq:triplet-superpotential}).  

One sees that the mixing between neutral fermions (gauginos, Higgsinos
and neutrinos) has a rich structure, with several off-diagonal RPV
entries which, as we show next, will induce masses (and mixings) for 
the neutrinos.

\section{Phenomenology}
\label{sec:phenomenology}

In what follows we will work in an approximation where we neglect the
radiatively induced neutrino masses with respect to those induced at
the tree level due to the presence of the triplet. We have checked
that there is a natural range of parameters in the model where the
use of this approximation is justified. For a thorough discussion of
loop-induced neutrino masses in BRpV see
refs.~\cite{Diaz:2003as,Hirsch:2000ef}.

\subsection{Neutrino masses and mixing angles}
\label{sec:neutr-mass-mixing}

One can provide an approximate analytical understanding of the tree
level neutrino masses and mixing angles, by using the previous form of
$\mathbf{M_N}$. This form is especially convenient because the
various sub-blocks in eq.~(\ref{eq:neutral-fermion-mass-matrix}) can
easily have vastly different orders of magnitude and one expects from
eq.~(\ref{eq:VEVs}) and eq.~(\ref{eq:tadpole-equations})

\begin{equation}
\mathbf{M_{\nu }},\mathbf{m_{\nu\Delta}},\mathbf{m_{\chi^0\Delta}}
\ll \mathbf{m_{BRpV}} \ll \mathbf{{\cal M}_{\chi^0}},\mathbf{M_{\Delta}}
\label{Hier}
\end{equation}
if $\epsilon_i < \mu$. More technically, in the limit where 
$\xi_{ij} \ll 1$, where $\xi= {\bf m_{BRpV}}\times {\cal M}_{\chi^0}^{-1}$ 
one can write the effective contribution of the BRpV parameters to the 
($3\times 3$) neutrino mass matrix as
 
\begin{equation}         
\mathbf{M_{\nu}^{eff}}=\mathbf{M_{\nu}} +\mathbf{M^{eff}_{Bilinear}}
\label{Meff}
\end{equation}
where the contributions from BRpV to the  neutrino mass matrix is given 
by~\cite{Hirsch:2000ef}

\begin{equation}
(\mathbf{M^{eff}_{Bilinear}})_{ij}=
\frac{M_1g^2+M_2g^{\prime2}}{4\det(\mathbf{\mathcal{M}_{\chi^0}})}
\Lambda_i\Lambda_j
\label{MBeff}
\end{equation}
with
\begin{equation}
\Lambda_i=\mu v_i+\epsilon_iv_d.
\end{equation}
With the observation eq.~(\ref{Hier}) one can further simplify the 
eigenvalue problem of eq.~(\ref{Meff}), by means of a perturbative 
diagonalization. Consider first eq.~(\ref{MBeff}). As has been 
discussed several times in the literature~\cite{Diaz:2003as}, this 
mass matrix is diagonalized by only two angles, 

\begin{equation}
\label{tetachooz}
\tan\theta_{13} = - \frac{\Lambda_e}
                   {(\Lambda_{\mu}^2+\Lambda_{\tau}^2)^{\frac{1}{2}}},
\end{equation}
\begin{equation}
\label{tetatm}
\tan\theta_{23} = - \frac{\Lambda_{\mu}}{\Lambda_{\tau}}.
\end{equation}
and leads to only one non-zero eigenvalue, given by 
\begin{equation}
m_{\nu}^{BRpV}=
\frac{M_1g^2+M_2g^{\prime2}}{4\det(\mathbf{\mathcal{M}_{\chi^0}})}
|\Lambda|^2,
\label{mTree}
\end{equation}
where $|\Lambda|^2 = \sum \Lambda_i^2$ 
The remaining angle $\theta_{12}$ is not defined in the bilinear 
only model at tree-level. The complete diagonalization of eq.~(\ref{Meff}), 
$\mathbf{\widehat{M}_{\nu }^{eff}}=
\mathbf{R}\cdot\mathbf{M_{\nu }^{eff}}\cdot\mathbf{R}^T$, where the 
matrix of eigenvectors $\mathbf{R}$ can be expressed as a product of three 
Euler rotations, can therefore be written as,

\begin{equation}
\begin{array}{ccl}
\mathbf{\widehat{M}_{\nu }^{effective}} & = &
\mathbf{R}\cdot\mathbf{M^{eff}_{Bilinear}}\cdot\mathbf{R}^T
+\mathbf{R}\cdot\mathbf{M_{\nu}}\cdot\mathbf{R}^T\\
& = & \mathbf{R_{12}}\cdot\mathbf{M'^{eff}_{Bilinear}}\cdot\mathbf{R_{12}}^T+
\mathbf{R_{12}}\cdot\mathbf{M'_{\nu}}\cdot\mathbf{R_{12}}^T
\end{array}\end{equation}
Since $\mathbf{M'^{eff}_{Bilinear}}$ is already diagonal, the solar 
angle is defined by the entries of $\mathbf{M'_{\nu}}$ only,
\begin{equation}\label{solang}
\tan(2\theta_{SOL})\simeq\frac{2(\mathbf{M'_{\nu}})_{12}}{(\mathbf{M'_{\nu}})_{11}-(\mathbf{M'_{\nu}})_{22}}
\end{equation}
Note that $\mathbf{M'_{\nu}}$ is not exactly diagonal after applying 
the rotation eq.~(\ref{solang}). However if 
$\mathbf{M'_{\nu}} \ll \mathbf{M'^{eff}_{Bilinear}}$, this small 
off-diagonals will change the angles defined in  eq.~(\ref{tetachooz}) 
and eq.~(\ref{tetatm}) only by a negligible amount.
Using the experimentally measured values of $\tan^2\theta_{ATM} \simeq 1$ 
and $\sin^22\theta_{\chooz} \ll 1$ one can finally find a simple formula for the 
solar angle ($\theta_{12}$) which is approximately given by
\begin{equation}
\tan(2\theta_{SOL})\simeq\frac{-2\sqrt{2}(h_{12} - h_{13})}
{-2h_{11}+h_{22}+h_{33} -2h_{23}}\equiv x
\label{simpang}
\end{equation}

From the above discussion one expects that, for reasonable ranges of
parameters, atmospheric neutrino physics is determined by the
bilinear parameters $\Lambda_i$, whereas the solar neutrino mass scale
depends mostly on the Yukawa couplings and the triplet mass. 

\begin{figure}[htbp]
\begin{center}
\includegraphics[width=75mm,height=50mm]{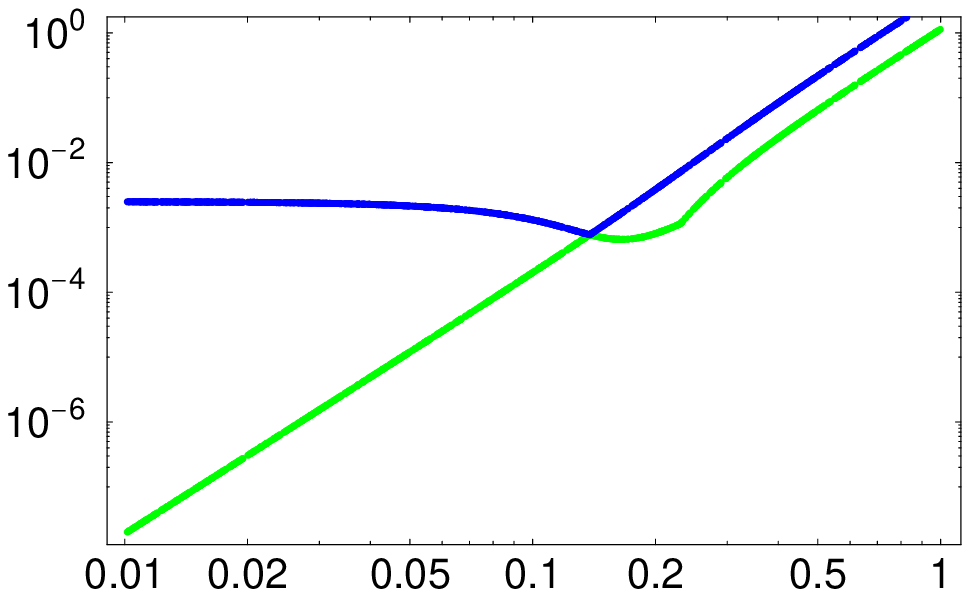}
\end{center}
\vskip-17mm
\hskip27mm
\begin{rotate}{90}
$\Dma,\;\Dms\;[eV^2]$
\end{rotate}
\vskip5mm
\hskip100mm
$h$
\vskip-50mm
\hskip50mm
$\Dma$
\vskip43.5mm
\vskip-28mm
\hskip60mm
$\Dms$
\vskip21.5mm

\caption{Typical behavior of $\Dma$ and $\Dms$ 
  vs. the total Higgs triplet Yukawa coupling $h$, for the 
  arbitrary choice of 
  $h_{11}=h_{23}=h/14$, $\,h_{22}=h_{33}=h_{13}=2\,h/7$, $\,h_{12}=0$.
  The triplet mass has been fixed at $M_{\Delta}=500$ GeV and the MSSM and
  BRpV choices are specified in the text.}
\label{msqh500}
\end{figure}

Fig.~\ref{msqh500} shows an example of the solar and atmospheric 
neutrino mass--squared differences 
as a function of the Yukawa parameter $h$ for a fixed value of the
triplet mass $M_{\Delta} = 500$ GeV. These results correspond to the
following choice of the MSSM parameters, $M_2 = 120$ GeV, $\mu = 500
\textrm{GeV}$, $\tan \beta = 5$, $A = -500 \textrm{GeV}$. In order to
ensure a) negligible loop corrections due to the bilinear parameters and 
b) correct neutrino mixing angles we have chosen the BRpV
parameters as follows: $\epsilon_1=\epsilon_2=\epsilon_3$ with
$\epsilon^2 = 10^{-3}|\vec{\Lambda}|$ and
$\Lambda_{\mu}=\Lambda_{\tau}=10 \Lambda_e$.  The absolute value of 
$|\vec{\Lambda}|$ can be estimated by 
$|\vec{\Lambda}|^2 = \frac{\sqrt{\Dma}}{\left(\frac{M_1g^2+M_2g'^2}{4
      \det{\mathcal{M}_{\chi^0}}}\right)}$. In the numerical estimate
we took the best fit $\Dma = 2.5\times 10^{-3} \textrm{eV}^2$ given
in~\cite{Pakvasa:2003zv}.

One sees that, for the hierarchical spectra produced by the model,
$m_{\nu_2} \simeq \sqrt{\Dms}$ scales approximately like
$m_{\nu_2} \sim h^2$. This is expected from 
eq.~(\ref{eq:neutrino-mass-matrix}) and eq.~(\ref{eq:VEVs}). For values
of $h \simeq {\cal O}(0.1)$, $\Dms$ is in the range of the
LMA-MSW solution of the solar neutrino problem. For larger values of
$M_{\Delta}$ the resulting $m_{\nu_2}$ gets smaller approximately like
$1/M_{\Delta}$. From Fig.~\ref{msqh500} one also sees that for large 
values of the Yukawas both solar and atmospheric masses are generated 
by the triplet.

To check to which degree the simplified results for the solar angle 
discussed above hold we have
constructed a set of randomly chosen sample points and diagonalized
the neutrino-neutralino mass matrix numerically.
Points were chosen as follows. For the MSSM parameters we scan 
randomly over the following ranges: $m_0$ in the interval [$0,1$] 
TeV, $M_2$ and $\mu$ from [$0,500$] GeV, with both signs for $\mu$, 
and $\tan\beta$ in [$2,15$]. The resulting SUSY spectra were checked 
to obey existing lower limits on sparticle searches. For the BRpV 
parameters consistency with the atmospheric neutrino data requires 
$|\Lambda|$ in the range [$0.05,0.15$] GeV$^2$, $\Lambda_{\mu} 
\simeq \Lambda_{\tau}$ and $\Lambda_{e} \le 0.3 
\sqrt{\Lambda_{\mu}^2+\Lambda_{\tau}^2}$. 

To reduce the number of free parameters we assume
$M_T\equiv\mu_{\Delta}=M_{\Delta_u}=M_{\Delta_d}=B_{\Delta}$.  We then have
calculated neutrino masses and mixing angles for several values of the
triplet mass, scanning randomly over the Yukawa couplings with the
over-restrictive constraint $h \equiv \sum_{i \le j} h_{ij} \le 1$.

\begin{figure}[htbp]
\begin{center}
\includegraphics[width=75mm,height=50mm]{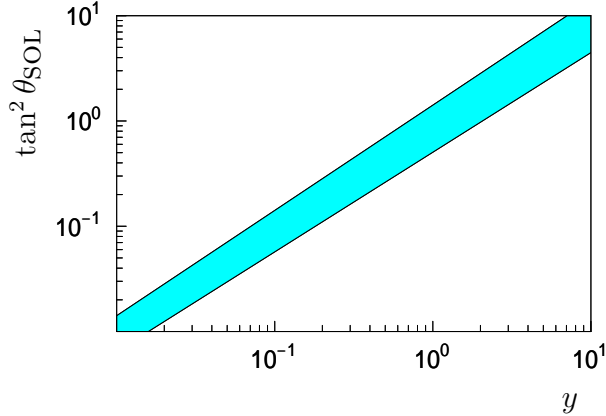}
\end{center}
\vskip-40mm
\hskip29mm
\begin{rotate}{90}
$\tan^2 \theta_\Sol$
\end{rotate}
\vskip27mm
\hskip99mm
$y$
\caption{Solar neutrino mixing angle vs. the ratio of triplet Yukawa couplings
  $y\equiv\tan^2\left(\frac{\arctan(x)}{2}\right)$, where
  $x\equiv\frac{-2\sqrt{2}(h_{12}-h_{13})}{-2h_{11}+h_{22}+h_{33}-2h_{23}}$
  and $M_T=500$ GeV.}
\label{solar-angle-yuk}
\end{figure}

Fig.~\ref{solar-angle-yuk} shows the numerically calculated
$\tan^2\theta_\Sol$ versus our simple formula eq.~(\ref{simpang}).
Clearly, for the region of interest the simple approximation works
surprisingly well. Note, that deviations between the exact and the 
approximate results mainly occur in the region of parameter space 
where $h_{12} \simeq h_{13}$.

\subsection{Implications for Accelerators}
\label{sec:impl-accel}

Since in our model R-parity is violated, the lightest supersymmetric 
particle will decay. As has been 
shown in \cite{Hirsch:2002ys,Restrepo:2001me,Porod:2000hv}, 
bilinear parameters can then be traced through the study of 
LSP decays. This feature will also remain to be true in the current 
model. We will not repeat the discussion and instead concentrate 
on the Higgs triplet in the following.

\begin{figure}[htbp]
\begin{center}
\includegraphics[width=75mm,height=50mm]{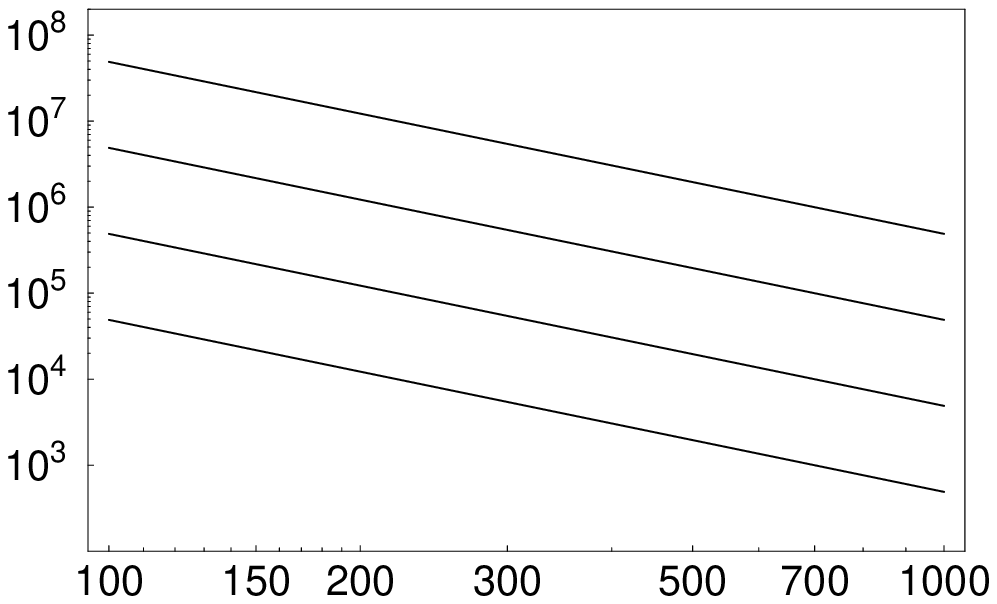}
\end{center}

\vskip-28mm
\hskip27mm
\begin{rotate}{90}
$\sigma(\sqrt{s}=M_{\Delta_u})$[fb]
\end{rotate}

\vskip16mm
\hskip85mm
$M_{\Delta_u}$[GeV]

\caption{Cross section for the production of the Higgs triplet at center of 
  mass energy equal to the triplet mass, $\sigma(\sqrt{s}=M_{\Delta_u})$, vs.
  the triplet mass. The four lines corresponds to a branching ratio (from 
  top to bottom) of 
  $BR(e^-e^-\to \Delta_u^{--})=10^{-1}$, $10^{-2}$, $10^{-3}$ and $10^{-4}$.}
\label{crossvsmt}
\end{figure}

One of the characteristic features of the triplet model of neutrino
mass is the presence of doubly charged Higgs bosons $\Delta^{--}_u$.
Here we consider its production cross section at an $e^- e^-$ linear
collider at 500 GeV center of mass energy\cite{Gunion:1998ii}. 
In Fig.~\ref{crossvsmt} we
present the s-channel production cross section for a doubly
charged Higgs boson as a function of its mass.
For typical expected luminosities of 500 inverse femtobarns per
year~\cite{Badelek:2001xb} this implies a very large number of events,
half a million or more, for a 500 GeV mass, depending on the leptonic
branching ratio in question.

The next issue are the decays of such Higgs bosons.  Here we come to
the most remarkable feature of the present model, namely that the
decays of the doubly charged Higgs bosons are a perfect tracer of the
neutrino mixing angles. The situation here is similar to that found in
the simplest bilinear R-parity model of neutrino mass considered in
refs.~\cite{Hirsch:2002ys,Restrepo:2001me,Porod:2000hv} (and
references therein). There it was found that, depending on the nature
of the lightest supersymmetric particle, its decays patterns reflect
in a simple way either the solar or the atmospheric mixing angles.
Here we have in addition that the doubly charged Higgs bosons decay
according to the solar mixing angle.

\begin{figure}[htbp]
\begin{center}
\includegraphics[width=75mm,height=50mm]{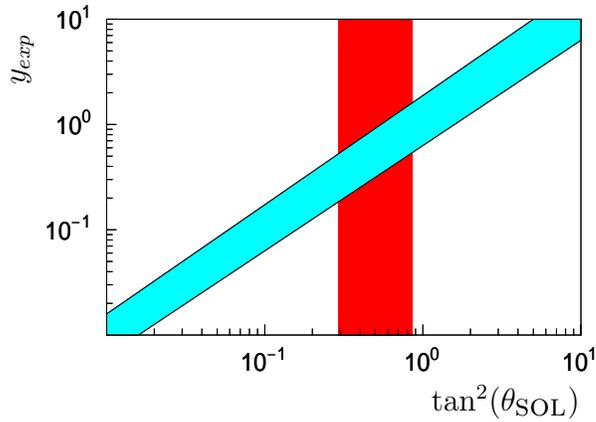}
\end{center}
\vskip-50mm
\hskip29mm
\begin{rotate}{90}
$y_{exp}$
\end{rotate}

\vskip37mm
\hskip83mm
$\tan^2(\theta_\Sol)$
\caption{Ratio of doubly charged Higgs boson decay branching ratios 
  indicated by the variable $y_{exp}$ of eqs.~(\ref{eq:yratio}) and
  (\ref{eq:xratio}) vs. the solar neutrino mixing angle. The
  vertical band indicates currently favored values (see
  ref.~\cite{Pakvasa:2003zv} and references therein).}
\label{tansqsolvsyexp01}
\end{figure}

In Fig.~\ref{tansqsolvsyexp01} we give the ratio of doubly charged
Higgs boson decay branching ratios versus the solar neutrino mixing
angle. The ratio of
doubly charged Higgs boson decay branching ratios we consider is
specified by the variable
\begin{equation}
  \label{eq:yratio}
y_{exp}\equiv\tan^2\left(\frac{\arctan(x_{exp})}{2}\right)  
\end{equation}
where
\begin{equation}
  \label{eq:xratio}
x_{exp}\equiv\frac{-2\sqrt{2}(\sqrt{BR_{12}}-\sqrt{BR_{13}})}{-2\sqrt{2 BR_{11}}+\sqrt{2 BR_{22}}+\sqrt{2 BR_{33}}-2\sqrt{BR_{23}}}
\end{equation}
with $BR_{ij}$ denoting the measured branching ratio for the process 
($\Delta_u^{--} \to l_i^-l_j^-$). Note that the band in the plot includes 
an assumed $10\%$ uncertainty in the measured branching ratios. The triplet
mass has been fixed at $M_{\Delta_u}=500$ GeV.
As can be seen from the figure, there is a very strong correlation
between the pattern of Higgs decays and the mixing angle involved in
the solar neutrino problem. The range permitted by current solar and
reactor neutrino data~\cite{Pakvasa:2003zv} is indicated by the vertical band in
Fig.~\ref{tansqsolvsyexp01}. This correlation can be used as the basis
for a reconstruction of neutrino angles using only accelerator
experiments. This provides a cross-check of the determination provided
by laboratory and underground searches for neutrino oscillations.

\section{Summary and Conclusions}
\label{sec:summary-conclusions}

We have extended the minimal supersymmetric standard model by adding
bilinear R-parity violation as well as a pair of Higgs triplet
superfields.  The neutral components of the Higgs triplets develop
small induced vacuum expectation values (VEVs) which depend
quadratically upon the bilinear R-parity breaking parameters. In this
scheme, for reasonable values of parameters, the atmospheric neutrino mass 
scale arises from bilinear
R-parity breaking while the solar neutrino mass scale is generated
from the small Higgs triplet VEVs.  We have calculated the pattern of
neutrino masses and mixing angles in this model and shown how the
model can be tested at future colliders. The branching ratios of the
doubly charged triplet decays are related to the solar neutrino angle
via a simple formula. Similarly the atmospheric mixing can be inferred
from the neutralino decay branching ratios, as discussed in
ref.~\cite{Porod:2000hv}. This will allow a full reconstruction of
neutrino angles purely from high energy accelerator experiments.
The model will be tested in a straightforward way should a high
luminosity and center-of-mass energy linear collider ever be built.

\section*{Acknowledgments}

We thank E.J. Chun, M.A. D\'{\i}az and J. Rom\~ao for useful discussions. 
This work was supported by Spanish grant BFM2002-00345, by the European 
Commission RTN grant HPRN-CT-2000-00148 and a bilateral CSIC-COLCIENCIAS 
agreement.  M. H. is supported by a Spanish MCyT Ramon y Cajal contract.  
A. V. was supported by a fellowship from Generalitat Valenciana.

\end{document}